\documentclass[sigconf]{acmart}
\AtBeginDocument{%
  \providecommand\BibTeX{{%
    Bib\TeX}}}

\copyrightyear{2026}
\acmYear{2026}
\setcopyright{cc}
\setcctype{by}
\acmConference[EASE '26]{International Conference on Evaluation and Assessment in Software Engineering}{June 09--12, 2026}{Glasgow, United Kingdom}
\acmBooktitle{International Conference on Evaluation and Assessment in Software Engineering (EASE '26), June 09--12, 2026, Glasgow, United Kingdom}
\acmDOI{10.1145/3816483.3816594}
\acmISBN{979-8-4007-2348-3/2026/06}

\usepackage{algorithmic}
\usepackage{graphicx}
\usepackage{textcomp}
\usepackage{xcolor}
\def\BibTeX{{\rm B\kern-.05em{\sc i\kern-.025em b}\kern-.08em
    T\kern-.1667em\lower.7ex\hbox{E}\kern-.125emX}}

\usepackage{listings}
\usepackage{adjustbox}

\usepackage{booktabs}
\usepackage{multirow}

\usepackage{framed}

\usepackage{float}
\restylefloat{table}

\usepackage[most]{tcolorbox}

\lstdefinestyle{prompt}{
    language={},
    basicstyle=\ttfamily\small,
    backgroundcolor=\color{lightgray},
    frame=single,
    rulecolor=\color{darkgray},
    framesep=1pt, 
    xleftmargin=8pt,
    xrightmargin=8pt,
    aboveskip=5pt,
    belowskip=12pt,
    showstringspaces=false,
    breaklines=true,
    breakatwhitespace=true,
    tabsize=2,
    captionpos=b,
    numbers=none,
    keywordstyle=,
    stringstyle=,
    commentstyle=,
    identifierstyle=
}

\usepackage{enumitem}
\usepackage{tikz}
\usetikzlibrary{shadows,arrows.meta,positioning}
\usepackage{pgfplots}
\usepackage{pgfplotstable}
\usepgfplotslibrary{groupplots}
\usepgfplotslibrary{statistics}
\pgfplotsset{compat=1.18}
\usepackage{xurl}

\begin{document}

\title{Reassessing Dataset Quality and Fine-Tuning Practices for Vulnerability Detection on PrimeVul dataset}

\author{Norbert Sándor Szolnoki}
\authornote{Both authors contributed equally to this research.}
\email{szolnoki@inf.u-szeged.hu}
\orcid{0009-0000-6351-9000}
\affiliation{%
  \institution{University of Szeged}
  \city{Szeged}
  \country{Hungary}
}

\author{Gábor Antal}
\authornotemark[1]
\orcid{0000-0002-3002-8624}
\email{antal@inf.u-szeged.hu}
\affiliation{%
  \institution{University of Szeged}
  \city{Szeged}
  \country{Hungary}
}
\additionalaffiliation{%
  \institution{FrontEndART Ltd.}
  \country{Hungary}
}



\begin{abstract}
Recent studies report promising results from supervised fine-tuning of pre-trained code language models for automated vulnerability detection on benchmarks such as PrimeVul. In this paper, we present a large-scale empirical evaluation of both model performance and dataset quality in this setting.
We systematically fine-tune three widely used encoder models (CodeBERT, GraphCodeBERT, and UniXCoder) across five imbalance-mitigation strategies, extensive hyperparameter sweeps, and multiple random seeds, resulting in 770 independent training runs.

Our results reveal a consistent pattern. While standard metrics such as F1 and ROC-AUC occasionally suggest improvements, security-critical and calibration-aware measures tell a different story. Matthews Correlation Coefficient remains near zero, predicted probabilities are poorly calibrated, and evaluation under strict false-positive constraints reveals high false-negative rates. Neither imbalance handling, threshold optimization, nor post-hoc calibration recovers a meaningful discriminative signal.
We also conduct a data quality assessment of the PrimeVul benchmark and identify structural issues such as minimal semantic differences between vulnerable and fixed functions and annotation noise introduced by commit-level labeling.
These findings suggest that the PrimeVul dataset, in its current form, presents significant challenges for reliable AI-based vulnerability detection in software engineering with the methodologies employed in this study. Our study highlights the need for more robust dataset design, improved evaluation protocols, and a rethinking of fine-tuning practices for security-critical tasks.
\end{abstract}




\maketitle

\section{Introduction}

Vulnerability detection in source code is a central task in software security and has become a critical research frontier due to the increasing size, complexity, and exposure level of modern software systems. Traditional static analysis techniques frequently have high false positive rates and are difficult to scale across heterogeneous codebases~\cite{bavota2015large,li2018vulpecker}.
Recent transformer-based code models suggest that fine-tuning pre-trained language models on vulnerability datasets may allow them to learn patterns of insecure behavior and outperform traditional static analysis~\cite{feng2020codebert,guo2021graphcodebert,guo2022unixcoder}.
As a result, fine-tuning models such as CodeBERT, GraphCodeBERT, and UniXCoder has emerged as a promising approach in vulnerability classification research \cite{li2021vuldeepecker,semasaba2020literature}.

However, despite promising results in prior work, the effectiveness of these models in real-world security settings remains uncertain.
Given the increasing role of AI in software engineering, especially in security-critical domains, understanding the correlation between model behavior and dataset properties is essential.
Recent studies suggest that high reported performance may reflect dataset artifacts rather than generalizable learning. Examples include inconsistent labeling, duplicated functional patterns across train and test splits, and minimal semantic differences between vulnerable and fixed code versions \cite{chakraborty2021deep,chirkow2022empirical}. Moreover, commonly used metrics such as F1 score or ROC-AUC quantify statistical separability but fail to reflect operational security concerns such as false negatives under strict false-positive constraints \cite{prabhu2020reliability,ye2022misleading}. These observations raise broader concerns about both model reliability and dataset quality in SE-specific AI tasks.

This paper presents a systematic fine-tuning and diagnostic evaluation of CodeBERT, GraphCodeBERT, and UniXCoder on the PrimeVul dataset, a widely used benchmark for function-level vulnerability detection. Our study jointly assesses model performance and dataset quality, focusing on the real-world relevance, robustness, and calibration of the learned classifiers. Specifically, we investigate the following research questions:

\begin{itemize}
	\item \textbf{RQ1:} Does supervised fine-tuning on PrimeVul produce consistent improvements in standard metrics (F1, ROC-AUC) across models, seeds, and imbalance mitigation strategies?
	\item \textbf{RQ2:} Are these improvements robust and security-relevant when evaluated using calibration error (ECE), Matthews correlation coefficient (MCC), and VD-S (false negative rate under strict false positive constraints)?
	\item \textbf{RQ3:} Which characteristics of the PrimeVul dataset may help explain unstable or misleading model behavior under the fine-tuning setups evaluated in this study?
\end{itemize}

Our findings reveal that across more than one hundred training configurations, standard metrics occasionally suggest performance gains, while security-critical metrics indicate unreliable or near-random behavior. In several cases, ROC-AUC remains close to 0.5 despite high F1 scores, and models fail to generalize beyond trivial threshold adjustments. Furthermore, we identify multiple dataset characteristics that may help explain the observed learning difficulties, including near-identical function pairs and noise introduced by commit-level labeling heuristics.

Contrary to expectations from prior literature, our results show no evidence that existing encoder models can be reliably fine-tuned on PrimeVul for practical vulnerability detection. We provide a negative but actionable outcome that highlights limitations in both model pipelines and dataset construction. Our diagnostic evaluation contributes to a growing body of work emphasizing the need for rigorously designed benchmarks and evaluation protocols in software engineering AI.

\textbf{Contributions.}
This work presents a comprehensive empirical evaluation of three widely used pre-trained encoder models (CodeBERT, GraphCodeBERT, and UniXCoder) under supervised fine-tuning on the PrimeVul benchmark. We systematically evaluate five commonly adopted imbalance mitigation strategies and a broad range of hyperparameter configurations, resulting in a large-scale, multi-seed experimental study. Beyond standard performance metrics, we provide a calibration-aware and risk-aware analysis showing that metrics such as F1 and ROC-AUC may give misleading impressions of model effectiveness. Finally, we contribute a data quality assessment of PrimeVul, highlighting structural characteristics that may affect learnability and reliability. Under the experimental settings considered in this study, we find no evidence that the evaluated standard encoder-based fine-tuning pipelines yield dependable vulnerability detection performance on PrimeVul. These results motivate further investigation into dataset design choices and evaluation practices in security-focused SE tasks.

The remainder of the paper reviews relevant background and related work (Sections 2–3), details the experimental methodology (Section 4), presents results (Section 5), discusses implications and limitations (Sections 6–7), and concludes with directions for future research (Sections 8–9).

\section{Background}

\subsection{Vulnerability Detection in Source Code}
Software vulnerability detection aims to identify code fragments that
introduce security weaknesses exploitable by an adversary. Formally, given a
function or method $x$, the task is to learn a mapping
$f: x \rightarrow \{0,1\}$, where $1$ denotes a vulnerability. The task is
challenging due to incomplete specifications of what constitutes a
vulnerability, semantic diversity, and dependency on execution context
\cite{semasaba2020literature}.

\subsection{Transformer-based Code Models}
Recent advances in pretrained transformer encoders have shaped the direction
of static vulnerability detection. CodeBERT \cite{feng2020codebert},
GraphCodeBERT \cite{guo2021graphcodebert} and UniXcoder \cite{guo2022unixcoder}
extend masked language modelling with code-specific objectives such as
data-flow prediction or cross-modal alignment. These models are frequently
adopted as general-purpose feature extractors in downstream classification.
Their success on tasks such as code search, summarization and clone detection
has motivated their application to security tasks.

\subsection{The PrimeVul Benchmark}
PrimeVul \cite{ding2024primevul} is a function-level vulnerability detection
dataset designed to mitigate biases found in earlier resources. It employs
temporal splits, removes exact duplicates and aligns samples with documented
CVE and CWE identifiers. Importantly, PrimeVul introduces \textbf{risk-aware
metrics} such as VD-S to reflect deployment contexts where false positives
must remain below strict operational tolerances.

\subsection{Evaluation Metrics in Security Contexts}
Security classification is inherently imbalanced and cost-asymmetric.
Accordingly, traditional metrics such as accuracy and F1 may provide
misleading conclusions \cite{yao2021impact}. Matthews Correlation
Coefficient is recommended for imbalanced settings and has been shown to be
more stable \cite{chicco2020mcc}. Calibration metrics such as Expected
Calibration Error and Brier Score measure probabilistic reliability, which is
critical for operational decision-making \cite{naeini2015ece}. PrimeVul
further proposes VD-S to assess failure modes under strict false-positive
budget constraints. These metrics jointly provide a multi-perspective
evaluation complementary to the conventional F1 score.

\section{Related Work}

\subsection{Encoder Models for Vulnerability Detection}
Pretrained transformer encoders are widely deployed in vulnerability
classification. Studies based on CodeBERT \cite{feng2020codebert},
GraphCodeBERT \cite{guo2021graphcodebert} and UniXcoder
\cite{guo2022unixcoder} commonly report improvements on benchmarks such as
BigVul and Devign \cite{chakraborty2021deep}. However,
subsequent analyses indicate that these gains often fail to generalize
beyond benchmark-specific artifacts, including duplicated patterns,
spurious correlations, and dataset-specific biases
\cite{napier2023empirical}. Recent work explores architectural extensions or alternative learning objectives, such as heterogeneous source-level intermediate representations in HGVul \cite{song2022hgvul} or hierarchical supervised contrastive learning for CWE classification \cite{ji2024contrastive}, reporting performance
improvements under controlled experimental conditions.

\subsection{PrimeVul-based Studies}
PrimeVul has enabled research on vulnerability detection in more realistic
learning settings by introducing temporal splits, function-level labels, and risk-aware evaluation metrics.
Li et al.\ propose CleanVul, an
LLM-assisted data filtering pipeline designed to mitigate annotation noise
in commit-based datasets~\cite{li2024cleanvul}.
Wen et al.\ introduce a reinforcement-based curriculum learning approach using synthetic preference signals to improve encoder robustness~\cite{wen2025revd}.
Other approaches combine auxiliary semantic signals, such as commit metadata or documentation cues, with representation learning to improve detection performance.
These studies suggest that model behavior on PrimeVul is highly sensitive to dataset curation, auxiliary supervision, and training protocol choices.
Straightforward supervised fine-tuning of encoder models often
yields inconsistent or degraded results when compared to earlier
benchmarks.

\subsection{Evaluation and Metric Reliability}
Previous research on evaluation methodology indicates that model rankings may change when transitioning from accuracy or F1 score to correlation-based, calibration-aware, or risk-sensitive metrics \cite{mishra2021robust}. Wakili et al.\ demonstrate that calibration-aware evaluation is essential to avoid overestimating the operational utility of machine learning models in security monitoring systems \cite{chicco2020mcc}. Ding et al.\ formalize the VD-S metric to expose failure modes under strict false-positive constraints, reflecting realistic deployment settings \cite{ding2024primevul}. Across the literature, there is increasing recognition that evaluation protocols significantly shape reported findings, particularly in the presence of
dataset noise, class imbalance and temporal variance, which can lead to contradictory conclusions under standard fine-tuning setups.

\subsection{Methodological Differences in Prior PrimeVul Studies}
Many previous studies show positive outcomes using transformer-based models for PrimeVul. However, these outcomes are often achieved under experimental conditions that significantly differ from ours. Common methodological choices include reporting single-seed results, tuning decision thresholds on test data, focusing exclusively on accuracy or F1 score, or omitting security-relevant constraints such as low false positive rates. In addition, few studies analyze the impact of near-duplicate functions or paired vulnerable and fixed samples on learning stability. Our work complements these efforts by systematically evaluating robustness across seeds and configurations, incorporating calibration- and risk-aware metrics, and explicitly analyzing dataset characteristics that influence model behavior.

\section{Methodology}
\label{sec:methodology}

This section describes the experimental design used to evaluate
standard fine-tuning practices under controlled and reproducible
conditions.

\subsection{Problem Formulation}

We consider function-level vulnerability detection as a binary classification
task. Each instance is a function (or method) body $x$ extracted from a C or
C++ project, and the label $y \in \{0,1\}$ indicates whether $x$ contains a
vulnerability ($y = 1$) or not ($y = 0$). Given a training set
$\{(x_i, y_i)\}_{i=1}^N$, the goal is to learn a classifier
$f_\theta : x \mapsto p_\theta(y=1 \mid x)$ that outputs a probability
estimate for the vulnerable class. A decision threshold $\tau \in [0,1]$ is
then used to obtain a binary prediction
$\hat{y} = \mathbb{I}\big[p_\theta(y=1 \mid x) \ge \tau\big]$.

\subsection{Dataset and Preprocessing}

We use the PrimeVul dataset \cite{ding2024primevul} in its JSONL format,
which provides three disjoint splits: training, validation, and test. Each
entry contains at least:

\begin{itemize}
    \item \texttt{func}: the function-level source code,
    \item \texttt{target}: the binary vulnerability label,
    \item \texttt{commit\_message}: the commit message (when available).
\end{itemize}

We parse the JSONL files line by line using the Python standard library,
convert them into \texttt{pandas} DataFrames, and then into
\texttt{datasets.Dataset} objects. For each instance, we construct a single
input field \texttt{code}:

\begin{itemize}
    \item by default, \texttt{code} is set to the raw function text
    (\texttt{func});
    \item when commit context is enabled, \texttt{code} is the concatenation
    of \texttt{commit\_message}, two newline characters, and \texttt{func}.
\end{itemize}

The numeric label is stored in a field \texttt{labels}. We do not perform
any additional filtering, de-duplication or manual cleaning beyond the
original PrimeVul preparation. The validation split is used only for model
selection, threshold tuning, and calibration; the test split is held out for
final evaluation.

\subsection{Models and Tokenization}

We evaluate three widely used pre-trained transformer encoders for source
code:

\begin{itemize}
    \item CodeBERT (\texttt{microsoft/codebert-base}) \cite{feng2020codebert},
    \item GraphCodeBERT (\texttt{microsoft/graphcodebert-base}) \cite{guo2021graphcodebert},
    \item UniXcoder (\texttt{microsoft/unixcoder-base}) \cite{guo2022unixcoder}.
\end{itemize}

For each encoder, we instantiate a binary classifier using \newline
\texttt{AutoModelForSequenceClassification} with \texttt{num\_labels = 2}.
We employ the corresponding \texttt{AutoTokenizer} for each model. The
\texttt{code} field is tokenized with truncation and maximum-length padding.
We explore three maximum sequence lengths, \newline $L \in \{128, 256, 512\}$, which
determines the amount of code (and optional commit context) visible to the
model.

\subsection{Imbalance Handling Strategies}

PrimeVul exhibits an imbalanced class distribution with significantly fewer
vulnerable functions than benign functions. To study the effect of common
imbalance mitigation techniques, we define five training strategies:

\begin{description}
    \item[Baseline] The original training distribution is used without
 resampling, and the loss is the standard cross-entropy.

    \item[Class weights] We compute inverse-frequency class weights on the
    original training labels. Let $n_0$ and $n_1$ denote the number of benign
    and vulnerable examples, and $n = n_0 + n_1$. The weight for class $c$ is
    \[
      w_c = \frac{n}{2 n_c}, \quad c \in \{0,1\},
    \]
    and we use these weights in the cross-entropy loss.

    \item[Undersampling] All vulnerable samples are retained. Benign samples
    are randomly undersampled without replacement such that the final ratio is
    approximately $3:1$ (benign to vulnerable). The resulting training set is
    shuffled before each run.

    \item[Oversampling] All benign samples are retained. Vulnerable samples
    are oversampled with replacement until the vulnerable class accounts for
    approximately $25\%$ of the training set. When the original proportion of
    vulnerable samples already exceeds this value, no oversampling is applied.

    \item[Focal loss] The original training distribution is preserved, but
    the loss is replaced by focal loss with focusing parameter
    $\gamma = 2.0$. For a sample with true class $y$ and predicted
    probability $p_y$, the per-sample loss is
    \[
      \ell_{\text{focal}} = - (1 - p_y)^\gamma \log(p_y + \varepsilon),
    \]
    where $\varepsilon$ is a small constant to avoid numerical issues.
\end{description}

Resampling is applied only to the training split; the validation and test splits
remain unchanged.

\subsection{Training Procedure}

We used the Hugging Face \texttt{Trainer} API. A custom trainer which publicly available, \texttt{ImbalanceTrainer}, extends the base class and overrides the \texttt{compute\_loss} method to select between unweighted cross-entropy, class-weighted cross-entropy and focal loss based on the chosen strategy. Apart from this loss selection, the training loop follows the standard implementation.

For each model and strategy we perform a small hyperparameter grid search
over:

\begin{itemize}
    \item learning rate $lr \in \{1\cdot10^{-5}, 2\cdot10^{-5}, 3\cdot10^{-5}\}$,
    \item maximum sequence length $L \in \{128, 256, 512\}$,
    \item training batch size $B_{\text{train}} \in \{8, 16\}$.
\end{itemize}

The evaluation batch size is fixed to $B_{\text{eval}} = 16$ and the number
of epochs is fixed to $E = 5$. Each configuration is run with three random seeds
$s \in \{1,2,3\}$ using \texttt{set\_seed}. We enable mixed-precision
training on GPUs (fp16) in all experiments. We do not apply early
stopping. We do not retain intermediate checkpoints in order to reduce
storage and, the model is trained for the full five epochs and evaluated once at
the end. The hyperparameters above were set as static variables to evaluate all the models with the same settings and make our results reproducible. We have not tested what the best hyperparameters are, and it is not the scope of our research.

A combination of model, strategy, hyperparameters and seed defines a
single run. For each run we store all metrics and predictions in a dedicated
directory that encodes the configuration.

\subsection{Calibration and Threshold Selection}

We evaluate each trained model both before and after post-hoc calibration.

\paragraph{Uncalibrated evaluation.}
We first obtain logits for the validation and test sets using the trained
model and convert them to class probabilities.
We retain $p_\theta(y=1 \mid x)$ as the score for the vulnerable class.

On the validation set we perform a one-dimensional search over decision
thresholds $\tau \in \{0.01, 0.02, \ldots, 0.99\}$ and select the threshold
$\tau^\star$ that maximizes the validation F1 score. We then compute all
metrics on both validation and test sets at (i) the default threshold
$\tau = 0.5$ and (ii) the validation-optimal threshold $\tau^\star$.

\paragraph{Temperature scaling.}
To improve probability calibration, we apply temperature scaling on the
validation set. We introduce a scalar temperature parameter $T > 0$ and
define scaled logits as $z / T$. The value of $T$ is estimated by minimizing
the negative log-likelihood on the validation set using L-BFGS (up to 50
iterations). The calibrated model is then applied to validation and test
logits, and probabilities are recomputed.

We repeat the threshold search on calibrated validation probabilities and
obtain a calibrated optimal threshold $\tilde{\tau}^\star$. As before, we
compute all metrics at the default threshold $\tau = 0.5$ and at
$\tilde{\tau}^\star$ for both the validation and test sets.

\subsection{Evaluation of Metrics}

Let $y$ denote the vector of true labels and $p$ the predicted
probabilities for the vulnerable class. For a fixed decision threshold
$\tau$, we derive binary predictions $\hat{y}$ and compute the following
metrics:

\begin{itemize}
	\item accuracy and balanced accuracy,
	\item F1 score,
	\item Matthews Correlation Coefficient (MCC) \cite{chirkow2022empirical}, which measures the
	correlation between predicted and true labels and remains informative
	under class imbalance; values near zero indicate performance
	statistically indistinguishable from random guessing,
	\item ROC-AUC\cite{fawcett2006introduction}, treating $p$ as continuous scores and measuring
	threshold-independent separability,
	\item Brier Score \cite{gneiting2007strictly}, defined as the mean squared error between $p$ and
	$y$, which captures the overall quality of probabilistic predictions,
	\item Expected Calibration Error (ECE)\cite{guo2017calibration}, computed using 15 equal-width
	probability bins, which quantifies the mismatch between predicted
	confidence and empirical accuracy; lower values indicate better-calibrated
 probabilities,
    \item VD-S, a risk-aware metric defined as the false negative rate at
	the threshold that maximizes the true positive rate while enforcing a
	false positive rate below $0.5\%$ \cite{ding2024primevul}, reflecting
	deployment scenarios where false positives are tightly constrained
	and missed vulnerabilities represent the dominant failure mode.
\end{itemize}

All metrics are computed on both validation and test sets, using
uncalibrated and calibrated probabilities, and evaluated at the default
threshold $\tau = 0.5$ as well as at thresholds optimized on the
validation split.

\subsection{Experiment Pipeline}

\begin{figure}[!ht]
\centering
\begin{tikzpicture}[
    node distance=1.6cm,
    process/.style={rectangle, rounded corners, draw=black, fill=gray!10, drop shadow, text width=4cm, align=center, minimum height=1cm},
    arrow/.style={->, thick}
]

\node[process] (load) {PrimeVul JSONL loading};
\node[process, below of=load] (prep) {Tokenization \\ + optional commit context};
\node[process, below of=prep] (imb) {Imbalance strategy \\ (baseline };
\node[process, below of=imb] (train) {Model fine-tuning \\ };
\node[process, below of=train] (cal) {Temperature scaling \\ };
\node[process, below of=cal] (thr) {Threshold optimisation \\ };
\node[process, below of=thr] (eval) {Final evaluation \\ MCC, AUC, ECE, VD-S, etc.};

\draw[arrow] (load) -- (prep);
\draw[arrow] (prep) -- (imb);
\draw[arrow] (imb) -- (train);
\draw[arrow] (train) -- (cal);
\draw[arrow] (cal) -- (thr);
\draw[arrow] (thr) -- (eval);
\end{tikzpicture}
\caption{End-to-end experimental pipeline used in this study.}
\label{fig:pipeline}
\end{figure}
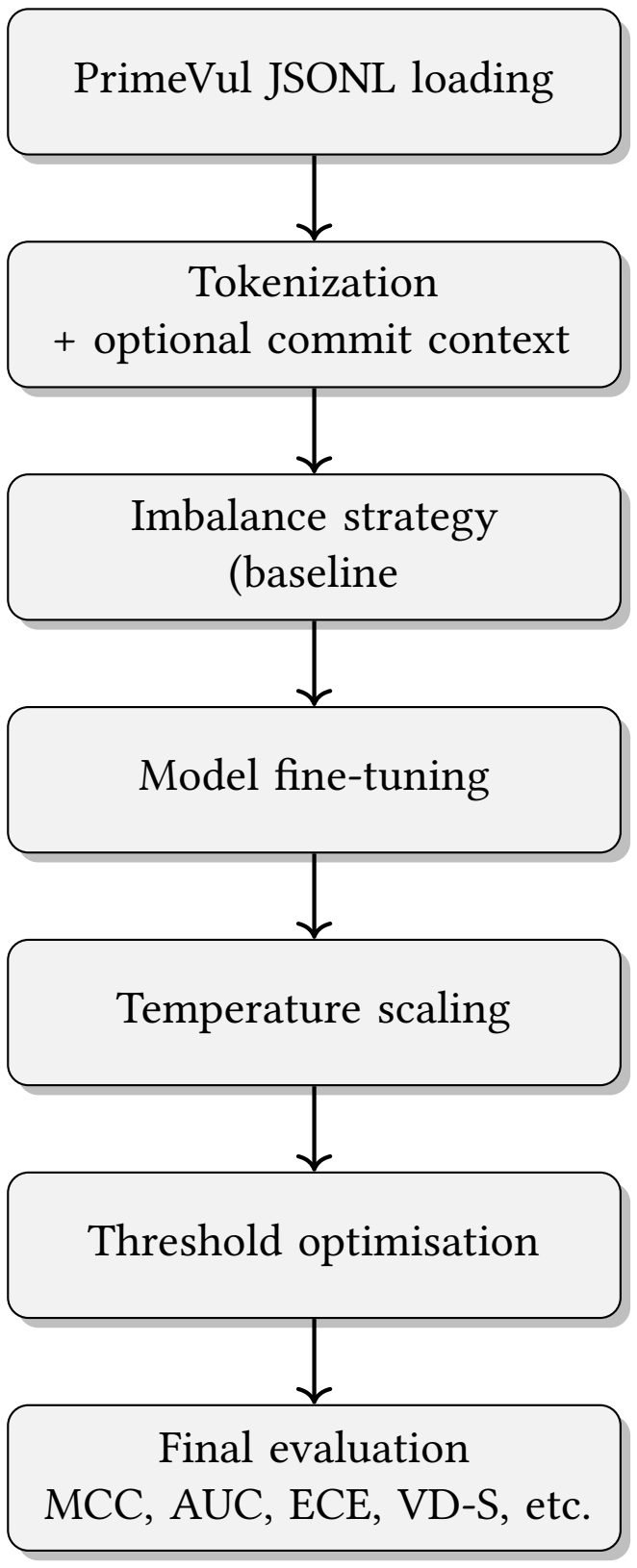

Figure~\ref{fig:pipeline} provides an overview of our experimental pipeline.
The pipeline is designed to trace the full lifecycle of a single experimental
configuration, from data ingestion to final metric reporting. Each experiment
is defined by the tuple
$(m, s, \ell, b, L, \gamma, r)$, where
$m$ is the pretrained model (CodeBERT, GraphCodeBERT, UniXcoder),
$s$ is the imbalance strategy (baseline, class weights, undersample,
oversample, focal), $\ell$ is the maximum sequence length,
$b$ is the batch size, $L$ is the learning rate, $\gamma$ is the focal loss parameter (if
applicable), and $r$ is the random seed.

\begin{enumerate}[label=\textbf{Step \arabic*.}]
  \item \textbf{Dataset loading.}
  We load the official PrimeVul JSONL files (train, validation, test).
  Fields \texttt{func} and \texttt{target} are used as inputs and labels.
  Optional commit-level context is prepended when \texttt{--use\_commit\_context} is enabled.

  \item \textbf{Preprocessing and tokenization.}
  Each function is tokenized using the model-specific tokenizer with fixed value of
  \texttt{max\_length} and \texttt{padding="max\_length"}.

  \item \textbf{Imbalance strategy application.}
  Depending on configuration, we apply one of:
  (i) no modification,
  (ii) class-weighted loss,
  (iii) undersampling of majority class,
  (iv) oversampling of minority class,
  (v) focal loss.
  Resampling is applied \emph{only} to the training split to avoid distribution shift.

  \item \textbf{Fine-tuning.}
  We fine-tune the model using the \texttt{Trainer} API
  with 5 epochs, AdamW optimizer, and the specified hyperparameters.
  Checkpointing is disabled to prevent accidental model selection and enforce
  strict comparability.

  \item \textbf{Prediction and calibration.}
  We evaluate the model on validation and test sets to obtain logits.
  We then fit a temperature scaling layer on the validation logits to produce
  calibrated probability estimates.

  \item \textbf{Threshold optimisation.}
  For both calibrated and uncalibrated predictions, we search over
  thresholds in $[0.01, 0.99]$ to maximise $F_{1}$ on the validation set.

  \item \textbf{Evaluation.}
  We compute the following metrics on the test set:
  accuracy, balanced accuracy, MCC, ROC-AUC, ECE, Brier score,
  and VD-S ($\mathrm{FNR}$ under $\mathrm{FPR} \leq 0.005$).
  All predictions, probabilities, thresholds, losses, and metric values
  are written to disk for post-hoc analysis.

  \item \textbf{Aggregation.}
  Results are aggregated across seeds.
  We report mean and standard deviation to capture model stability.
\end{enumerate}

\subsection{Design Rationale and Methodological Scope}

The experimental design of this study prioritizes breadth, control and
diagnostic coverage over architectural novelty. Rather than proposing
a new model or training objective, we intentionally adopt widely used
pretrained encoder architectures and standard fine-tuning pipelines
in order to evaluate a common experimental practice under controlled
and reproducible conditions. This choice allows observed performance
limitations to be attributed to task formulation, data properties and
evaluation protocol, rather than to idiosyncratic modeling decisions.

Hyperparameter ranges, imbalance mitigation strategies and calibration
procedures were selected to reflect established practice in the
vulnerability detection literature. At the same time, design choices
that could introduce optimistic bias, such as test-set threshold
tuning, early stopping on test metrics or selective reporting across
seeds were deliberately avoided. As a result, the study emphasizes
robustness and interpretability of outcomes over peak performance,
with the goal of providing a reliable empirical assessment of what
standard fine-tuning using the PrimeVul dataset.

\subsection{Data Quality Analysis of PrimeVul}
\label{sec:primevul-quality}

In addition to evaluating model performance, we conduct a diagnostic analysis of the PrimeVul dataset to examine dataset characteristics that may help explain the observed model behavior. Our results suggest that several structural characteristics of PrimeVul may limit its effectiveness as a training resource for vulnerability detection under the setups studied here.

First, we observe a high degree of lexical and syntactic similarity between paired vulnerable and fixed functions. Using TF-IDF vectorization over character-level $n$-grams ($n = 3$ to $5$), we compute cosine similarity between each vulnerable function and its corresponding fixed version across all paired instances in the dataset. Across 4,416 function pairs, the median similarity is 0.992 (IQR: 0.967–0.999), indicating near-identical lexical structure in most cases. Notably, 89.4\% of pairs exhibit similarity above 0.90 and 81.2\% exceed 0.95. These results indicate that vulnerable and fixed variants often differ only by minimal edits in the analyzed pairs (e.g., constant values, log messages, or localized line-level patches). Such extreme structural overlap may introduce substantial ambiguity into the learning task, because global sequence representations can be dominated by unchanged context rather than vulnerability-indicative features, potentially limiting separability for token-based encoder models.

Second, PrimeVul derives function-level binary labels from commits that fixes vulnerabilities. While this approach improves scalability and alignment with documented CVEs, it inherently relies on commit-level signals to infer fine-grained vulnerability semantics. As a result, individual functions modified in a fixing commit may include changes unrelated to the root vulnerability, incomplete fixes, or refactorings. Combined with high structural similarity between vulnerable and fixed versions and substantial class imbalance, this weak supervision setting may reduce the effective learning signal available to fine-tuned models.

Taken together, these findings highlight dataset characteristics that may help explain why standard supervised fine-tuning performs unreliably on PrimeVul in our experiments. The dataset’s minimal semantic variation between vulnerable and fixed functions, coarse-grained function-level labeling, and reliance on commit-derived annotations may weaken the effective supervision signal available to models. These characteristics make it difficult for encoder-based classifiers to learn robust and interpretable decision boundaries. We therefore advocate for improved annotation granularity, context-aware labeling, and change-level supervision to better support reliable AI-based vulnerability detection.

\paragraph{Experimental Sufficiency.}
To ensure that our findings are not an artifact of insufficient tuning, we conducted an extensive and systematic hyperparameter exploration. Across all experiments, we varied pretrained model architectures, random seeds, imbalance mitigation strategies, learning rates, sequence lengths, batch sizes, calibration settings, and decision thresholds. In total, this resulted in over 770 independent training runs and corresponding evaluations, all performed using fixed dataset splits and controlled execution environments.
The selected hyperparameter ranges (learning rates in the order of $10^{-5}$, batch sizes of 8–16, sequence lengths up to 512 tokens, and 3–5 training epochs) follow the standard fine-tuning regime established for BERT-style transformer models \cite{devlin2019bert} and widely adopted in subsequent studies on fine-tuning stability and code representation models \cite{mosbach2020stability,feng2020codebert,guo2021graphcodebert}. Further expansion of this search space would risk overfitting to PrimeVul rather than improving generalizable performance. We therefore consider the experimental coverage sufficient to draw reliable conclusions.

\section{Results}

We report results for 770 fine-tuning configurations spanning three
pretrained models (CodeBERT, GraphCodeBERT and UniXCoder), five imbalance
mitigation strategies, multiple hyperparameter settings and three random
seeds (Section~4). All test results use calibrated probabilities and
decision thresholds selected exclusively on the validation split. In our analysis, MCC, calibration error and VD-S are treated as
primary indicators of security-relevant performance, while accuracy,
F1 and ROC-AUC are reported for completeness.

\begin{table*}[t]
\centering
\caption{PrimeVul test performance for the best configuration of each model and imbalance strategy. 
Values are averaged over three seeds. 
F1, MCC and AUC are threshold-optimised on the validation set using calibrated probabilities. 
VD-S FNR is the false negative rate at the best achievable true positive rate under $\mathrm{FPR} \leq 0.005$.}
\label{tab:main_results}
\small
\begin{tabular}{llrrrrrrr}
\toprule
\textbf{Model} & \textbf{Strategy} & \textbf{F1} & \textbf{MCC} & \textbf{AUC} & \textbf{Brier} & \textbf{ECE} & \textbf{VD-S FNR} & \textbf{Thr.} \\
\midrule
codebert      & baseline      & 0.666 & -0.004 & 0.511 & 0.250 & 0.010 & 0.990 & 0.170 \\
codebert      & class\_weights & 0.667 &  0.000 & 0.510 & 0.250 & 0.009 & 0.992 & 0.010 \\
codebert      & focal         & 0.667 &  0.000 & 0.514 & 0.250 & 0.008 & 0.994 & 0.010 \\
codebert      & oversample    & 0.667 &  0.000 & 0.517 & 0.250 & 0.008 & 0.992 & 0.010 \\
codebert      & undersample   & 0.666 &  0.000 & 0.521 & 0.250 & 0.014 & 0.992 & 0.157 \\
\midrule
graphcodebert & baseline      & 0.667 &  0.000 & 0.498 & 0.250 & 0.002 & 0.997 & 0.170 \\
graphcodebert & class\_weights & 0.666 & -0.005 & 0.523 & 0.250 & 0.019 & 0.993 & 0.310 \\
graphcodebert & focal         & 0.667 &  0.000 & 0.499 & 0.250 & 0.004 & 0.996 & 0.010 \\
graphcodebert & oversample    & 0.665 &  0.009 & 0.523 & 0.250 & 0.014 & 0.994 & 0.167 \\
graphcodebert & undersample   & 0.667 &  0.017 & 0.522 & 0.250 & 0.016 & 0.992 & 0.130 \\
\midrule
unixcoder     & baseline      & 0.665 &  0.035 & 0.544 & 0.249 & 0.036 & 0.993 & 0.433 \\
unixcoder     & class\_weights & 0.665 &  0.021 & 0.542 & 0.249 & 0.024 & 0.991 & 0.437 \\
unixcoder     & focal         & 0.666 &  0.004 & 0.524 & 0.250 & 0.014 & 0.995 & 0.330 \\
unixcoder     & oversample    & 0.666 &  0.018 & 0.542 & 0.249 & 0.029 & 0.992 & 0.433 \\
unixcoder     & undersample   & 0.664 &  0.024 & 0.540 & 0.249 & 0.026 & 0.993 & 0.313 \\
\bottomrule
\end{tabular}
\end{table*}

\subsection{Overall Performance Across Models and Strategies}

Table~\ref{tab:main_results} reports the best-performing configuration
for each (model, imbalance strategy) pair, selected by validation MCC.
Across all evaluated settings, performance remains close to
random-guessing baselines. UniXCoder achieves the highest ROC-AUC values,
reaching up to 0.549, while CodeBERT and GraphCodeBERT generally remain
near or below 0.52. However, these differences do not translate into
meaningful improvements in MCC or balanced accuracy, which remain close
to zero across all models.

\begin{figure}[H]
	\centering
	\begin{tikzpicture}
		\begin{axis}[
			ybar,
			width=\linewidth,
			height=6cm,
			ymin=0.48,
			ymax=0.56,
			bar width=6pt,
			enlarge x limits=0.10,
			legend style={at={(0.5,1.02)},anchor=south,legend columns=3},
			ylabel={ROC-AUC},
			xtick=data,
			xticklabel style={rotate=45,anchor=east},
			symbolic x coords={
				CB-baseline, CB-cw, CB-focal, CB-over, CB-under,
				GCB-baseline, GCB-cw, GCB-focal, GCB-over, GCB-under,
				UX-baseline, UX-cw, UX-focal, UX-over, UX-under
			},
			nodes near coords,
			nodes near coords style={
				font=\scriptsize,
				/pgf/number format/fixed,
				/pgf/number format/precision=3
			},
			]
			\addplot coordinates {
				(CB-baseline,0.511028)
				(CB-cw,0.510364)
				(CB-focal,0.514174)
				(CB-over,0.517278)
				(CB-under,0.520949)
				(GCB-baseline,0.498492)
				(GCB-cw,0.522834)
				(GCB-focal,0.498917)
				(GCB-over,0.523396)
				(GCB-under,0.522353)
				(UX-baseline,0.543875)
				(UX-cw,0.542209)
				(UX-focal,0.523506)
				(UX-over,0.541793)
				(UX-under,0.540307)
			};
		\end{axis}
	\end{tikzpicture}
	\caption{Test ROC-AUC for each model and imbalance strategy, using the best hyperparameter configuration selected on validation MCC (after calibration and threshold optimisation).}
	\label{fig:auc_by_model_strategy}
\end{figure}
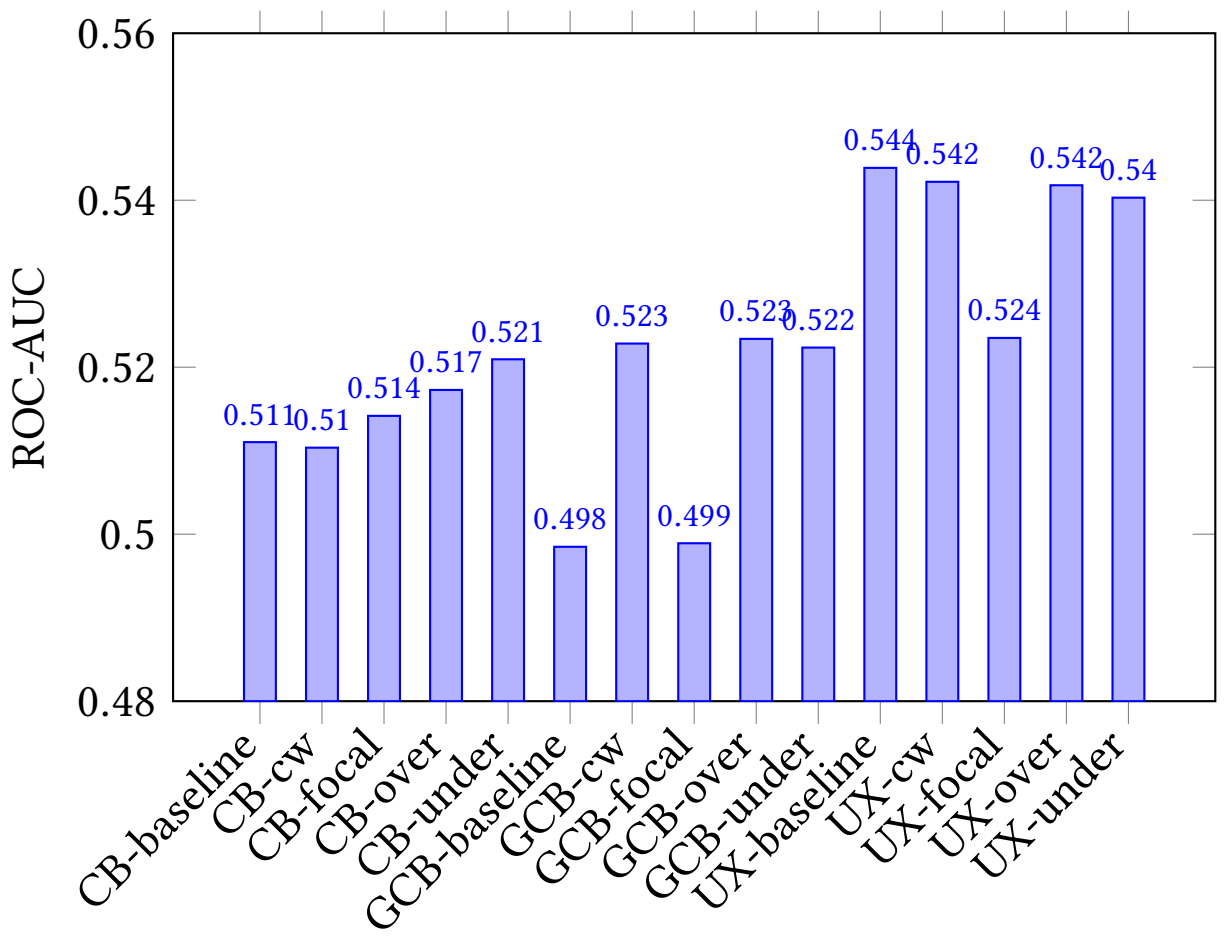

Figure~\ref{fig:auc_by_model_strategy} provides an overview of ROC-AUC
across imbalance mitigation strategies. No strategy consistently
outperforms the unmodified baseline across architectures. Observed
differences are small and model-dependent, indicating a lack of
systematic benefit from class weighting, oversampling, undersampling or
focal loss.

\subsection{Robustness and Hyperparameter Sensitivity}

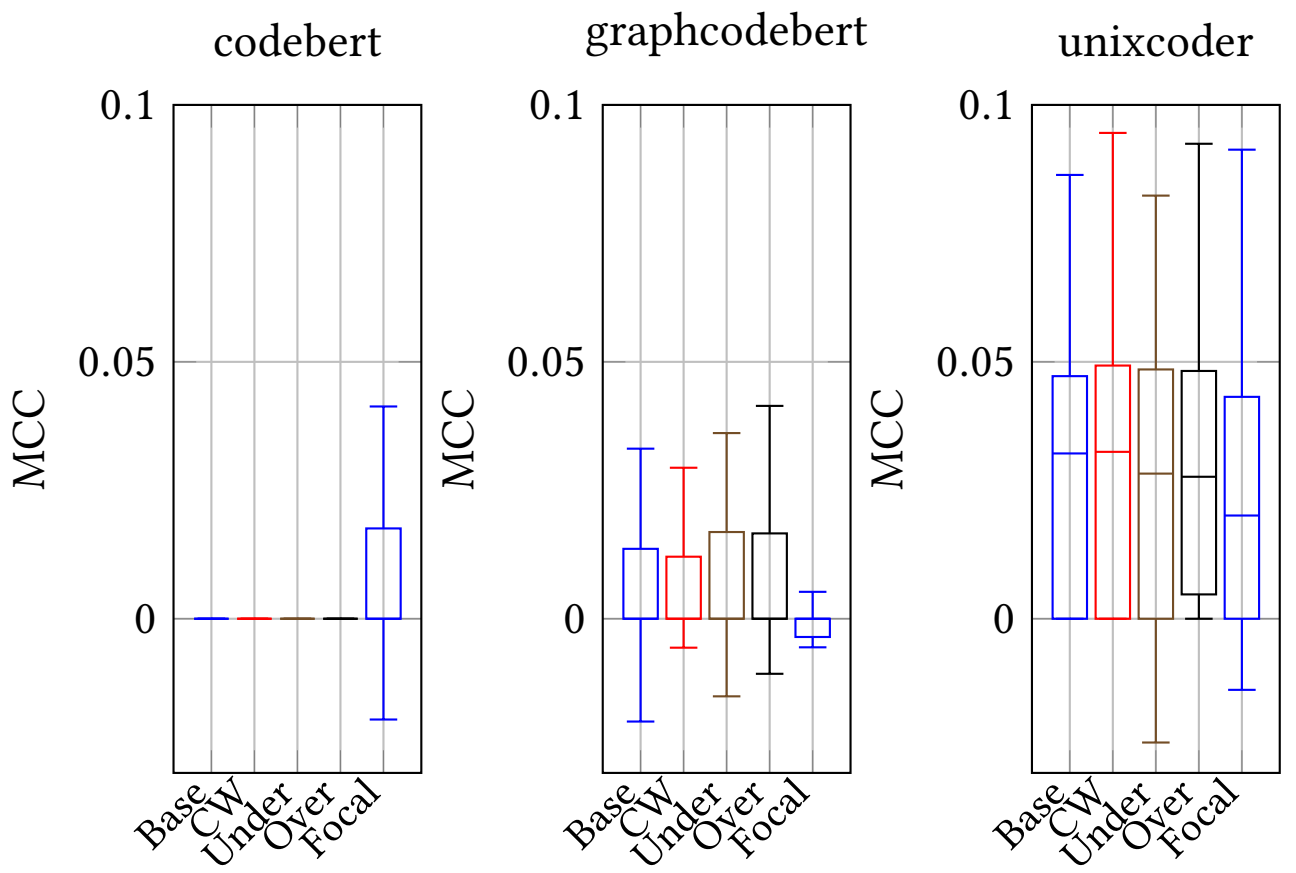
\begin{figure}[t]
	\centering
	\begin{tikzpicture}
		\begin{groupplot}[
			group style={group size=3 by 1, horizontal sep=1.2cm},
			width=0.38\linewidth,
			height=6cm,
			ylabel={MCC},
			xtick={1,2,3,4,5},
			xticklabels={Base,CW,Under,Over,Focal},
			x tick label style={rotate=45, anchor=east, font=\small},
			grid=major,
			scaled y ticks=false,
			yticklabel style={/pgf/number format/fixed},
			yticklabel style={/pgf/number format/precision=2},
			boxplot/draw direction=y,
			ymin=-0.03, ymax=0.10, 
			]
			
			\nextgroupplot[title={codebert}]
			\addplot+[boxplot prepared={draw position=1,
				lower whisker=0, lower quartile=0, median=0, upper quartile=0, upper whisker=0
			}] coordinates {};
			\addplot+[boxplot prepared={draw position=2,
				lower whisker=0, lower quartile=0, median=0, upper quartile=0, upper whisker=0
			}] coordinates {};
			\addplot+[boxplot prepared={draw position=3,
				lower whisker=0, lower quartile=0, median=0, upper quartile=0, upper whisker=0
			}] coordinates {};
			\addplot+[boxplot prepared={draw position=4,
				lower whisker=0, lower quartile=0, median=0, upper quartile=0, upper whisker=0
			}] coordinates {};
			\addplot+[boxplot prepared={draw position=5,
				lower whisker=-0.019608, lower quartile=0, median=0, upper quartile=0.017582, upper whisker=0.041315
			}] coordinates {};
			
			\nextgroupplot[title={graphcodebert}]
			\addplot+[boxplot prepared={draw position=1,
				lower whisker=-0.020001, lower quartile=0, median=0, upper quartile=0.013609, upper whisker=0.033103
			}] coordinates {};
			\addplot+[boxplot prepared={draw position=2,
				lower whisker=-0.005629, lower quartile=0, median=0, upper quartile=0.012079, upper whisker=0.029390
			}] coordinates {};
			\addplot+[boxplot prepared={draw position=3,
				lower whisker=-0.015097, lower quartile=0, median=0, upper quartile=0.016880, upper whisker=0.036138
			}] coordinates {};
			\addplot+[boxplot prepared={draw position=4,
				lower whisker=-0.010710, lower quartile=0, median=0, upper quartile=0.016614, upper whisker=0.041428
			}] coordinates {};
			\addplot+[boxplot prepared={draw position=5,
				lower whisker=-0.005555, lower quartile=-0.003545, median=0, upper quartile=0, upper whisker=0.005242
			}] coordinates {};
			
			\nextgroupplot[title={unixcoder}]
			\addplot+[boxplot prepared={draw position=1,
				lower whisker=0, lower quartile=0, median=0.032175, upper quartile=0.047218, upper whisker=0.086376
			}] coordinates {};
			\addplot+[boxplot prepared={draw position=2,
				lower whisker=0, lower quartile=0, median=0.032485, upper quartile=0.049278, upper whisker=0.094554
			}] coordinates {};
			\addplot+[boxplot prepared={draw position=3,
				lower whisker=-0.024084, lower quartile=0, median=0.028250, upper quartile=0.048521, upper whisker=0.082352
			}] coordinates {};
			\addplot+[boxplot prepared={draw position=4,
				lower whisker=0, lower quartile=0.004741, median=0.027635, upper quartile=0.048228, upper whisker=0.092442
			}] coordinates {};
			\addplot+[boxplot prepared={draw position=5,
				lower whisker=-0.013834, lower quartile=0, median=0.020093, upper quartile=0.043187, upper whisker=0.091292
			}] coordinates {};
			
		\end{groupplot}
	\end{tikzpicture}
	\caption{MCC across imbalance strategies and models. Distributions are computed across all seeds and hyperparameter combinations.}
	\label{fig:fig_box_mcc}
\end{figure}

To assess robustness, we analyze the distribution of performance across all hyperparameter configurations and random seeds. Figure~\ref{fig:fig_box_mcc} shows the distribution of test MCC values for each model. For all three architectures, MCC values are narrowly concentrated around zero, with limited variance across configurations. UniXCoder exhibits a slightly higher central tendency, but the distributions remain highly overlapping.

For CodeBERT, MCC remains effectively zero across all imbalance strategies and hyperparameter settings. Despite moderate F1 scores, this indicates predictions that are statistically indistinguishable from random guessing. Inspection of predicted probabilities reveals a strong concentration around 0.5, suggesting that the model fails to learn meaningful decision boundaries and instead converges toward majority-class behavior under class imbalance.

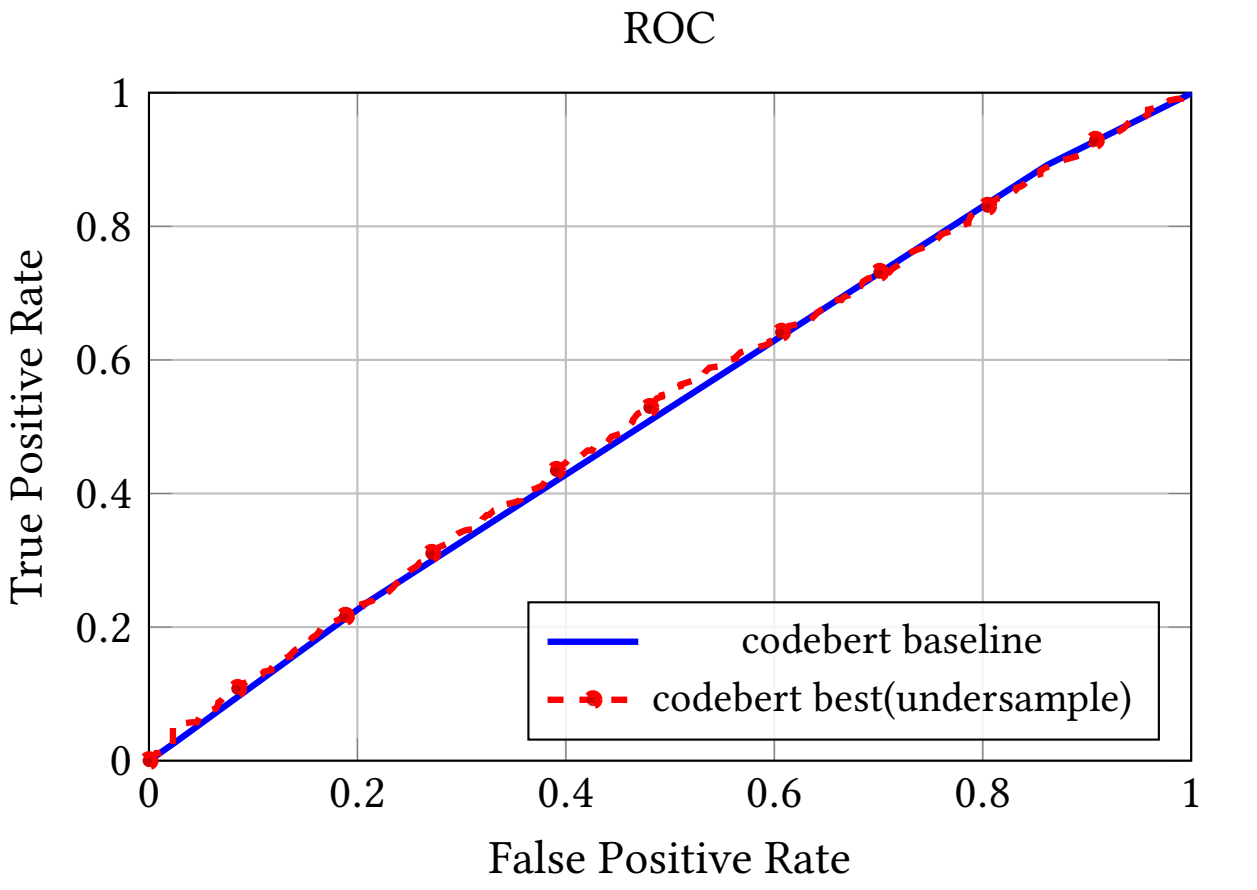
\begin{figure}[t]
\centering
\begin{tikzpicture}
\begin{axis}[
  width=\linewidth,
  height=6cm,
  xlabel={False Positive Rate}, ylabel={True Positive Rate},
  xmin=0, xmax=1.0, ymin=0, ymax=1,
  grid=major,
  legend pos=south east,
  legend style={fill opacity=0.85, draw opacity=1, text opacity=1, font=\small},
  title={ROC}
]
\addplot+[very thick, solid, mark=none] coordinates {(0.000000,0.000000) (0.025287,0.027586) (0.211494,0.239080) (0.862069,0.891954) (0.974713,0.979310) (0.995402,0.995402) (1.000000,1.000000)};
\addlegendentry{codebert baseline}
\addplot+[very thick, dashed, mark=*, mark repeat=25, mark size=1.2pt] coordinates {(0.000000,0.000000) (0.000000,0.002299) (0.004598,0.006897) (0.006897,0.011494) (0.011494,0.011494) (0.016092,0.013793) (0.020690,0.020690) (0.022989,0.022989) (0.022989,0.043678) (0.022989,0.052874) (0.027586,0.055172) (0.041379,0.057471) (0.045977,0.057471) (0.048276,0.064368) (0.050575,0.066667) (0.052874,0.071264) (0.055172,0.073563) (0.059770,0.075862) (0.064368,0.078161) (0.066667,0.085057) (0.066667,0.087356) (0.068966,0.089655) (0.071264,0.094253) (0.082759,0.098851) (0.082759,0.101149) (0.085057,0.108046) (0.087356,0.112644) (0.101149,0.124138) (0.103448,0.126437) (0.108046,0.128736) (0.110345,0.133333) (0.117241,0.135632) (0.117241,0.137931) (0.121839,0.147126) (0.131034,0.154023) (0.135632,0.156322) (0.135632,0.158621) (0.140230,0.165517) (0.142529,0.167816) (0.147126,0.172414) (0.149425,0.174713) (0.154023,0.177011) (0.156322,0.183908) (0.160920,0.188506) (0.163218,0.193103) (0.165517,0.200000) (0.170115,0.202299) (0.172414,0.204598) (0.174713,0.206897) (0.179310,0.209195) (0.188506,0.216092) (0.190805,0.216092) (0.197701,0.220690) (0.200000,0.222989) (0.204598,0.229885) (0.204598,0.234483) (0.209195,0.236782) (0.213793,0.239080) (0.216092,0.241379) (0.216092,0.243678) (0.220690,0.245977) (0.222989,0.248276) (0.227586,0.250575) (0.229885,0.252874) (0.234483,0.259770) (0.236782,0.264368) (0.241379,0.271264) (0.241379,0.275862) (0.245977,0.278161) (0.250575,0.285057) (0.252874,0.287356) (0.257471,0.291954) (0.259770,0.296552) (0.268966,0.301149) (0.268966,0.305747) (0.271264,0.310345) (0.278161,0.319540) (0.285057,0.324138) (0.285057,0.326437) (0.296552,0.340230) (0.303448,0.344828) (0.312644,0.347126) (0.312644,0.349425) (0.317241,0.351724) (0.319540,0.360920) (0.324138,0.367816) (0.326437,0.367816) (0.331034,0.374713) (0.335632,0.377011) (0.342529,0.379310) (0.342529,0.383908) (0.349425,0.386207) (0.354023,0.388506) (0.356322,0.390805) (0.356322,0.397701) (0.365517,0.404598) (0.370115,0.406897) (0.377011,0.411494) (0.379310,0.416092) (0.383908,0.429885) (0.390805,0.434483) (0.393103,0.439080) (0.397701,0.443678) (0.400000,0.445977) (0.402299,0.450575) (0.404598,0.452874) (0.406897,0.452874) (0.413793,0.455172) (0.420690,0.464368) (0.425287,0.464368) (0.427586,0.471264) (0.439080,0.473563) (0.441379,0.478161) (0.441379,0.480460) (0.443678,0.485057) (0.452874,0.489655) (0.455172,0.494253) (0.455172,0.496552) (0.459770,0.498851) (0.462069,0.503448) (0.464368,0.508046) (0.464368,0.510345) (0.466667,0.514943) (0.468966,0.519540) (0.478161,0.526437) (0.480460,0.528736) (0.485057,0.537931) (0.487356,0.542529) (0.491954,0.544828) (0.491954,0.547126) (0.503448,0.551724) (0.505747,0.558621) (0.510345,0.560920) (0.510345,0.563218) (0.524138,0.570115) (0.526437,0.572414) (0.528736,0.574713) (0.531034,0.574713) (0.533333,0.581609) (0.537931,0.588506) (0.549425,0.590805) (0.554023,0.595402) (0.556322,0.600000) (0.563218,0.602299) (0.563218,0.604598) (0.567816,0.611494) (0.593103,0.622989) (0.595402,0.625287) (0.595402,0.627586) (0.604598,0.634483) (0.606897,0.641379) (0.609195,0.643678) (0.611494,0.643678) (0.613793,0.650575) (0.620690,0.652874) (0.625287,0.655172) (0.632184,0.655172) (0.634483,0.659770) (0.639080,0.666667) (0.641379,0.671264) (0.648276,0.678161) (0.650575,0.685057) (0.652874,0.687356) (0.657471,0.689655) (0.664368,0.689655) (0.666667,0.694253) (0.671264,0.696552) (0.673563,0.701149) (0.673563,0.705747) (0.678161,0.708046) (0.682759,0.710345) (0.685057,0.717241) (0.689655,0.721839) (0.696552,0.724138) (0.698851,0.726437) (0.701149,0.731034) (0.710345,0.731034) (0.712644,0.735632) (0.714943,0.737931) (0.721839,0.744828) (0.724138,0.749425) (0.726437,0.751724) (0.728736,0.758621) (0.731034,0.760920) (0.737931,0.765517) (0.742529,0.767816) (0.747126,0.770115) (0.747126,0.772414) (0.754023,0.777011) (0.760920,0.788506) (0.770115,0.793103) (0.770115,0.795402) (0.777011,0.800000) (0.781609,0.802299) (0.786207,0.804598) (0.786207,0.806897) (0.788506,0.816092) (0.790805,0.818391) (0.793103,0.820690) (0.795402,0.822989) (0.804598,0.829885) (0.809195,0.832184) (0.811494,0.839080) (0.813793,0.841379) (0.825287,0.845977) (0.827586,0.848276) (0.832184,0.855172) (0.832184,0.857471) (0.836782,0.859770) (0.839080,0.862069) (0.843678,0.868966) (0.843678,0.873563) (0.850575,0.875862) (0.855172,0.878161) (0.855172,0.882759) (0.859770,0.887356) (0.864368,0.889655) (0.868966,0.894253) (0.871264,0.896552) (0.887356,0.903448) (0.891954,0.905747) (0.898851,0.917241) (0.901149,0.917241) (0.903448,0.919540) (0.905747,0.926437) (0.908046,0.928736) (0.910345,0.928736) (0.919540,0.933333) (0.924138,0.935632) (0.926437,0.937931) (0.928736,0.940230) (0.935632,0.947126) (0.937931,0.951724) (0.942529,0.958621) (0.947126,0.958621) (0.954023,0.960920) (0.956322,0.965517) (0.958621,0.967816) (0.958621,0.974713) (0.965517,0.977011) (0.970115,0.979310) (0.972414,0.981609) (0.972414,0.983908) (0.977011,0.986207) (0.979310,0.988506) (0.986207,0.990805) (0.990805,0.990805) (0.993103,0.995402) (0.995402,0.997701) (1.000000,1.000000)};
\addlegendentry{codebert best(undersample)}
\end{axis}
\end{tikzpicture}
\caption{ROC curves for representative runs. Substantial overlap indicates limited discriminative improvement across imbalance mitigation strategies and hyperparameter sweeps.}
\label{fig:fig_roc_full_codebert}
\end{figure}

Representative ROC curves further illustrate this behavior.
Figures~\ref{fig:fig_roc_full_codebert} and \ref{fig:fig_roc_full_unixcoder} compare
baseline and best-performing configurations for CodeBERT and UniXCoder,
respectively. Substantial overlap across the full operating range
indicates limited discriminative improvement despite extensive
hyperparameter tuning.

\input{fig/fig_roc_full_unixcoder.tex}

\subsection{Calibration and Confidence Behaviour}

\begin{figure}[t]
\centering
\begin{tikzpicture}
\begin{axis}[
  width=\linewidth,
  height=6cm,
  xlabel={Mean predicted probability}, ylabel={Empirical accuracy},
  xmin=0, xmax=1, ymin=0, ymax=1,
  grid=major,
  legend pos=south east,
  title={Reliability diagram}
]
\addplot[densely dashed] coordinates {(0,0) (1,1)};
\addlegendentry{Perfect calibration}
\addplot+[mark=*] coordinates {(0.450117,0.567742) (0.497490,0.518195) (0.542145,0.607143)};
\addlegendentry{unixcoder uncal}
\addplot+[mark=*] coordinates {(0.450116,0.567742) (0.497490,0.518195) (0.542147,0.607143)};
\addlegendentry{unixcoder calibrated}
\end{axis}
\end{tikzpicture}
\caption{Reliability diagram (15 equal-width bins) for representative runs. Temperature scaling improves probability calibration only marginally, consistent with residual ECE.}
\label{fig:fig_reliability_unixcoder}
\end{figure}
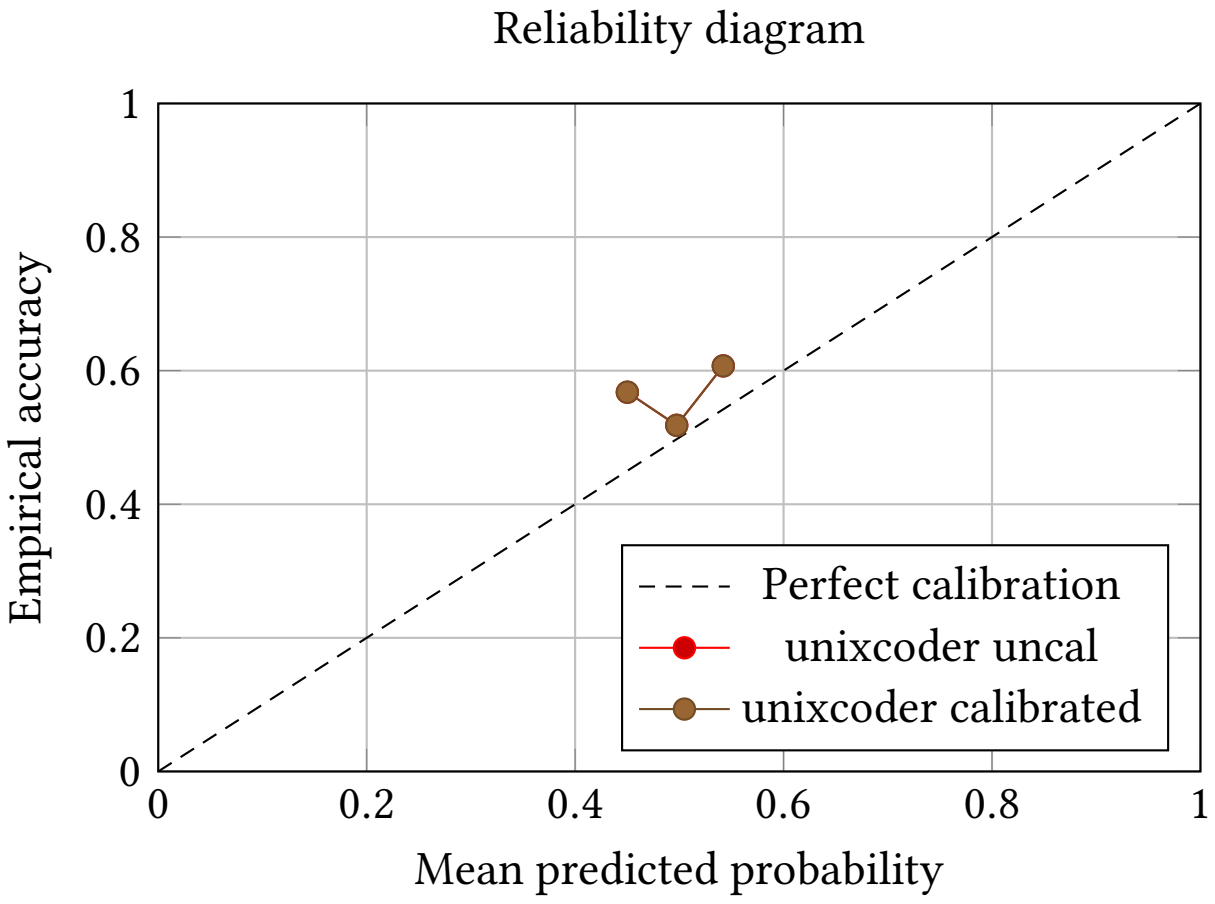

We next examine whether post-hoc calibration and decision threshold
optimization recover a useful signal. Figure~\ref{fig:fig_reliability_unixcoder}
shows the reliability diagram for the best UniXCoder configuration.
Temperature scaling slightly reduces expected calibration error, but
predicted probabilities remain tightly concentrated around 0.5.

Threshold optimization increases F1-scores on the validation split but
does not improve ROC-AUC, MCC or VD-S on the test set. Low–false-positive
operating points remain characterized by high false negative rates,
indicating that calibration and thresholding cannot compensate for the
lack of separability in the learned representations.

Taken together, these results are consistent with models relying on
shallow lexical or distributional cues rather than vulnerability-relevant
 semantic distinctions.

\section{Threats to Validity}

\subsection{Internal Validity}
The training pipeline contains several design choices that may impact the
measured outcomes. First, model performance reflects the behavior of the
default Hugging Face fine-tuning procedure with minor modifications to the loss
functions; alternative optimization schedules or regularization may lead to
different results. Second, the learning rate and sequence-length grids, while
covering commonly used values in prior work, are not exhaustive. It is
therefore possible that configurations outside our search space perform
better. Third, temperature scaling and threshold selection rely exclusively
on the validation split; if the validation set is not representative, the
calibrated results may be misaligned with deployment expectations.

To mitigate these threats, we fix random seeds, apply a systematic grid search
for all key hyperparameters, and store all intermediate outputs for
reproducibility. However, we cannot exclude the possibility that the observed
negative results are partially influenced by implementation-specific choices.

\subsection{External Validity}
Our conclusions are based solely on the PrimeVul dataset. While PrimeVul was
designed to reduce bias and improve ecological validity, it does not fully
capture the diversity of real-world development environments. In particular,
PrimeVul is dominated by C and C++ projects and focuses on function-level
granularity. Our findings may not generalize to other languages (e.g., Python,
JavaScript), other granularities (e.g., project-level or basic-block-level
analysis) or settings where dynamic behavior, build configurations, or
third-party dependencies influence vulnerability manifestation.

Furthermore, our study considers only three pretrained encoder models
(CodeBERT, GraphCodeBERT, UniXcoder). Transformer architectures with
specialized objectives (e.g., contrastive alignment, syntax-aware, or
graph-guided encoders) might exhibit different behavior.

\subsection{Construct Validity}
PrimeVul assigns binary labels per function, but vulnerability is ultimately a
semantic property that can depend on program state, execution context, or
usage patterns outside the function boundary. As a result, labels may not
always reflect the true presence or absence of a vulnerability in isolation.
Prior analyses indicate that PrimeVul contains mislabeled or ambiguous
instances due to commit-level annotation heuristics, incomplete fixing
patches, and functions that are syntactically distinct but semantically
related across commits \cite{ding2024primevul, li2024cleanvul}. Such label
noise may disproportionately harm models relying on local patterns, leading
to unstable predictions and calibration degradation.

To reduce construct bias, we do not introduce additional cleaning or
re-annotation that could alter dataset assumptions, and we report multiple
metrics (including MCC, calibration, and risk-aware VD-S) to avoid
over-interpreting single-score trends.

\section{Discussion}

\paragraph{Scope and Boundaries of the Study.}
This work does not claim that pretrained code language models are inherently unsuitable for vulnerability detection, nor that fine-tuning is ineffective in all settings. Rather, our findings are scoped to supervised whole-function binary classification on the PrimeVul dataset under standard fine-tuning protocols. Alternative formulations, such as change-level modeling, pairwise learning, or task-specific supervision, may yield different results. Our goal is to diagnose limitations in both the dataset and standard training pipeline, and to clarify their interplay in shaping reported outcomes.

\subsection{Implications for Dataset Design in AI for SE}

Our results suggest several data-related factors that may contribute to the limitations observed in encoder-based fine-tuning for vulnerability detection. Most notably, PrimeVul’s function pairs (vulnerable and fixed versions) are often nearly identical, differing only by small lexical edits such as a modified constant or a logging statement. As quantified in Section~4.10, this extreme structural overlap limits effective separability at the function level. When vulnerable and fixed variants share nearly identical lexical structure, global sequence representations are dominated by unchanged context rather than vulnerability-indicative features. This provides one plausible structural explanation for the near-zero MCC values observed across models in Section~5.

In addition, the use of commit-level labeling introduces significant annotation noise. Vulnerabilities are labeled based on changes committed together, which may include unrelated modifications or incomplete fixes. This weak supervision may misguide the model during training, especially in edge cases where vulnerable behavior is subtle or distributed across the program.

Taken together, these issues highlight the need for better dataset design practices in AI for software engineering. Specifically, future datasets should:
\begin{itemize}
	\item adopt vulnerability annotation strategies that isolate and highlight the actual source of insecure behavior,
	\item provide change-level or region-specific supervision rather than treating entire functions as binary-labeled units,
	\item encourage structural and semantic variation between positive and negative samples to support effective representation learning.
\end{itemize}

Datasets that reflect these principles are more likely to support robust, interpretable, and practically relevant AI models for security-focused tasks in SE.

\subsection{Implications for Evaluation Practice and Reproducibility}

Beyond data quality, our findings underscore the critical importance of evaluation protocol design in model assessment. Prior work often relies on single-seed experiments, fixed decision thresholds, and performance metrics that do not reflect operational requirements. In contrast, our large-scale multi-seed study reveals that seemingly strong results are frequently sensitive to threshold tuning and fail to generalize across configurations.

Specifically, we find that threshold optimization can inflate F1 scores without improving actual class separability, and that uncalibrated probability estimates yield poor reliability in low–false-positive regimes. Metrics such as MCC, ECE, and VD-S are more indicative of security-relevant behavior and should be prioritized in future work.

From a reproducibility standpoint, we recommend that future studies adopt evaluation practices that:
\begin{itemize}
	\item aggregate results across multiple seeds and configurations,
	\item report confidence intervals or standard deviations to indicate performance stability,
	\item include calibration- and risk-aware metrics that align with real-world constraints.
\end{itemize}

Such practices are essential for producing robust and verifiable claims about model effectiveness in SE contexts.

\subsection{Lessons Learned for Data-Centric AI in Software Security}

Our results suggest that the observed limitations of encoder fine-tuning on PrimeVul are consistent with both modeling limitations and data-related challenges. Apparent improvements under commonly used metrics such as F1 or ROC-AUC do not necessarily translate into security-relevant behavior, especially when label noise, class imbalance, and weak supervision dominate the learning signal.

Even with extensive hyperparameter tuning across more than 770 runs, the models consistently fail to generalize beyond trivial thresholds or baseline separability. This suggests that improvements to model architectures alone may be insufficient to address the core issue, namely that the learning task may be difficult under the current dataset and labeling formulation.

We conclude that progress in learning-based vulnerability detection will require a more integrated focus on dataset design, supervision quality, and evaluation rigor. In particular, aligning data representations with the semantics of vulnerability and the operational realities of secure software development is critical to enabling reliable AI systems in SE.

\section{Conclusion}

This paper presented a large-scale empirical evaluation of supervised
fine-tuning of pretrained code language models for function-level
vulnerability detection on the PrimeVul benchmark. Across 770
configurations spanning three widely used encoder architectures,
multiple imbalance mitigation strategies and extensive hyperparameter
sweeps, we found no evidence that standard fine-tuning yields reliable
or security-relevant performance improvements. While some conventional metrics suggested improvement, these gains did
not persist under calibration-aware or risk-sensitive evaluation.

Taken together, our results suggest that commonly reported performance
improvements on PrimeVul should be interpreted with caution. In
particular, strong results under single metrics or fixed thresholds
may not reflect practical effectiveness in security-critical settings.
By applying rigorous evaluation protocols and deployment-oriented
metrics, this study clarifies the current limitations of baseline
encoder fine-tuning on realistic vulnerability benchmarks.

We hope that these findings help redirect future research toward
problem formulations, datasets and learning objectives that are more
closely aligned with the semantic and operational characteristics of
real-world software vulnerabilities. 

In particular, our findings highlight several dataset-level issues with high lexical and semantic similarity between fixed and vulnerable variants, noise introduced by function-level labels, and limited diversity of vulnerability patterns—that hinder effective model training. These insights underscore the need for more semantically rich, precisely annotated, and representative datasets to advance AI-based vulnerability detection in realistic SE settings.

\section{Data Availability}

All data and code required to reproduce the experiments reported in this
study are publicly available via Zenodo \cite{anonimus2026primevul}.
The archive includes the processed PrimeVul splits used in our
experiments, training and evaluation scripts, configuration files, and
scripts for metric computation and result aggregation. To support
reproducibility, we also provide all random seeds and hyperparameter
settings used in the study.

\section{Future Work}

Our findings suggest multiple directions for future research.

\paragraph{Architectural Adaptations.}
A natural extension is to evaluate architectures that incorporate structural
priors beyond token sequences. Examples include models with explicit data-flow
representations, code property graphs, or syntax-guided attention mechanisms.
Several recent studies indicate that combining pretrained encoders with
contrastive or graph-based objectives may mitigate noise sensitivity
\cite{ji2024contrastive}, and systematic evaluation of such
hybrid approaches on PrimeVul remains open.

\paragraph{Siamese and Pair-aware Training.}
As PrimeVul contains semantically linked pairs of vulnerable and fixed code
instances originating from the same commit, future work could leverage this
implicit pairing with Siamese or triplet networks. These approaches could
directly model the relationship between vulnerable and non-vulnerable variants
of the same function, potentially improving discrimination under label noise
and representation drift.

\paragraph{Data Quality and Re-annotation.}
Our results highlight the influence of dataset noise and inconsistent
annotation granularity. Future efforts could investigate semi-automated
re-annotation workflows, such as LLM-assisted validation or human-in-the-loop
correction, building on recent work in dataset cleaning for vulnerability
detection \cite{li2024cleanvul}. A cleaner and more semantically coherent
dataset would permit more fine-grained conclusions about architectural
contributions.

\paragraph{Robust Training Regimes.}
Several training strategies remain unexplored in this context. Promising
directions include curriculum learning that orders samples by estimated
difficulty, noise-robust loss functions, uncertainty-aware optimization or
test-time adaptation for out-of-distribution robustness. Moreover, hard
negative mining or active learning may help address the imbalance between
benign and vulnerable instances more effectively than static resampling.

\paragraph{Evaluation Protocols.}
While this work focuses on function-level classification, future research
could consider multi-level evaluation across files, classes or projects, and
integration with contextual signals such as build configuration or dependency
graphs. Additionally, adapting risk-aware metrics to deployment-specific
tolerances (e.g., alternative VD-S thresholds or cost-sensitive metrics) may
provide further insight into practical applicability.

\paragraph{Benchmark Diversity.}
Our conclusions are tied to PrimeVul. Extending the study to other benchmarks
(e.g., BigVul, Devign, Vul4J or CleanVul) could help determine whether the
observed instability is specific to PrimeVul’s properties or a broader
limitation of baseline encoder fine-tuning. Cross-dataset evaluation would
also clarify the transferability of calibration and imbalance strategies.

\section{Acknowledgment}
This work was supported in part by the European Union through the European Cybersecurity Competence Centre under Grant no. 101128082 (APPTake).

\bibliographystyle{ACM-Reference-Format}
\bibliography{references}

\end{document}